\documentclass[onecolumn,showpacs,preprintnumbers,amsmath,amssymb,prb]{revtex4}

\usepackage{graphicx}
\usepackage{dcolumn}
\usepackage{bm}
\usepackage{color}

\begin{document}
\title{Phase and amplitude of Aharonov-Bohm oscillations in nonlinear three-terminal transport through a double quantum dot}

\author{Toshihiro Kubo$^{1}$}
 \email{kubo.toshihiro@lab.ntt.co.jp}
\author{Yuki Ichigo$^{1,2}$}
\author{Yasuhiro Tokura$^{1,2}$}%
 \affiliation{%
$^1$NTT Basic Research Laboratories, NTT Corporation, Atsugi-shi, Kanagawa 243-0198, Japan\\
$^2$Department of Physics, Tokyo University of Science, Shinjuku-ku, Tokyo 162-8601, Japan
}%

\date{\today}
             
\begin{abstract}
We study three-terminal linear and nonlinear transport through an Aharonov-Bohm interferometer containing a double quantum dot using the nonequilibrium Green's function method. Under the condition that one of the three terminals is a voltage probe, we show that the linear conductance is symmetric with respect to the magnetic field (phase symmetry). However, in the nonlinear transport regime, the phase symmetry is broken. Unlike two-terminal transport, the phase symmetry is broken even in noninteracting electron systems. Based on the lowest-order nonlinear conductance coefficient with respect to the source-drain bias voltage, we discuss the direction in which the phase shifts with the magnetic field. When the higher harmonic components of the Aharonov-Bohm oscillations are negligible, the phaseshift is a monotonically increasing function with respect to the source-drain bias voltage. To observe the Aharonov-Bohm oscillations with higher visibility, we need strong coupling between the quantum dots and the voltage probe. However, this leads to dephasing since the voltage probe acts as a B\"{u}ttiker dephasing probe. The interplay between such antithetic concepts provides a peak in the visibility of the Aharonov-Bohm oscillations when the coupling between the quantum dots and the voltage probe changes.
\end{abstract}

\pacs{73.23.-b, 73.63.Kv, 73.40.Gk, 05.60.Gg}

\maketitle
\section{Introduction}
The manifestation of quantum phase coherence forms one of the foundations of the physics of mesoscopic systems\cite{imry}, and is attracting the attention of many physicists. Quantum phase coherence is detectable by quantum interference experiments employing, for example, the Aharonov-Bohm (AB) effect\cite{ab}. In an AB interferometer containing a quantum dot (QD), the AB effects in the transport properties have been widely studied both theoretically\cite{abt1,abt2} and experimentally\cite{abe1,abe2,abe3,abe5}. The experiments show that phase coherence is maintained during the tunneling process through a QD.

By applying a finite bias voltage across a QD system, we can easily realize a nonequilibrium steady state condition. Therefore, QD systems provide a suitable stage for testing theories related to nonequilibirum systems. It is reasonable to expect that driving a system out of equilibrium will provide a new understanding of quantum interference effects. Some of symmetries present at equilibrium, which underline a linear response, may be broken, and at the same time new qualitative features may emerge.

The Onsager-Casimir symmetry relation states that the linear conductance in two-terminal systems should be symmetric with respect to an external magnetic field (phase rigidity)\cite{onsager,casimir}. In the nonlinear transport regime, however, it is not necessary for this phase symmetry to be satisfied. Recently, phase symmetry breaking in the nonlinear transport regime in two-terminal systems has been extensively studied both theoretically\cite{two-terminal1,two-terminal2,two-terminal3,two-terminal4} and experimentally\cite{two-terminal5,two-terminal6,two-terminal7,two-terminal8,two-terminal9,two-terminal10}. Phase rigidity is not enforced in a two-terminal conductor if the conductor is interacting with another subsystem in a nonequilibrium situation\cite{nonc1,nonc2,nonc3}. Moreover, in a multi-terminal conductor that includes the lossy channels, the phase symmetry breaks since the additional reservoirs allow losses of current and lead to the violation of unitarity \cite{multi1,multi2}. In Ref. \onlinecite{multi3}, B\"{u}ttiker had shown that the phase symmetry relation is satisfied in linear transport regime through multi-terminal device that all terminals except the source and drain reservoirs are voltage probes. In this paper, we consider a three-terminal system that satisfies unitarity where one of the three terminals is a voltage probe. The question is now, when the voltage probe is included in the AB interferometer, is the phase symmetry broken due to the voltage (electrochemical potential) fluctuation of the voltage probe in the nonlinear transport regime? The voltage probe is mathematically equivalent to the B\"{u}ttiker dephasing probe\cite{buttiker}. In this approach, a system is connected to a virtual electron reservoir through a fictitious voltage probe. Electrons are scattered into such a probe, lose their phase memories with a certain probability, and are then reinjected into the system. Thus, the voltage probe induces dephasing and at the same time it assists quantum phase coherence as part of the AB interferometer. Moreover, in two-terminal systems, the phase symmetry is not broken in noninteracting electron systems. A voltage probe is an infinite impedance terminal with zero net current and imposes a constraint. Then, we expect that the phase symmetry may be broken due to this constraint even in non-interacting electron systems.

In this paper, we study the \textit{phase} and \textit{amplitude} of the AB oscillation in a three-terminal AB interferometer device to address the following two main issues: (i) Is the phase symmetry broken in nonlinear transport through a three-terminal AB interferometer that includes a voltage probe? To clarify this, we investigate the phase of the AB oscillation in the lowest-order nonlinear conductance coefficient with respect to the bias voltage. (ii) The other major issue this paper addresses is the dephasing effects caused by the voltage probe, which is a component of an AB interferometer. We examine the amplitude of AB oscillations in transport properties to discuss the way in which the interference effect is suppressed by coupling with the voltage probe.

Here we investigate linear and nonlinear three-terminal transport through an AB interferometer containing a double quantum dot (DQD) using the nonequilibrium Green's function method \cite{schwinger,keldysh}. Recently, the AB effects in an AB interferometer containing a DQD have been thoroughly examined\cite{abt3,abt4,abt5,kubo,tokura,abe4,abe6,abe7}. In laterally coupled DQD systems, coherent indirect coupling between two QDs via a reservoir is essential in terms of coherent transport through a DQD\cite{kubo,tokura}. We introduce the coherent indirect coupling parameter $\alpha$, which characterizes the strength of the indirect coupling between two QDs via a reservoir. We consider three reservoirs, namely a source ($S$), a drain ($D$), and a voltage probe ($\varphi$) as shown in Fig. \ref{system}. The  electrochemical potential of the voltage probe is determined in order to satisfy the condition that the net current through the voltage probe vanishes. This is equivalent to a B\"{u}ttiker dephasing probe\cite{buttiker}. Thus, the coupling between the QDs and the voltage probe gives rise to the dephasing in the electronic states of the DQD. The coherent indirect coupling in a laterally coupled double quantum dot characterizes the coherence between two quantum dots. Thus, to study the interplay between the inter-dot coherence due to the coherent indirect coupling and the dephasing due to the voltage probe, the three-terminal system with a voltage probe coupled to a DQD is intersting. We show that the linear conductance is symmetric with respect to the magnetic field and show that this phase symmetry is broken in the nonlinear transport regime. In two-terminal systems, an electron-electron interaction is essential for breaking the phase symmetry\cite{two-terminal1,two-terminal2,two-terminal3}. However, in a three-terminal system including a voltage probe, we show that the phase symmetry is broken even in \textit{noninteracting} electron systems. When the higher harmonic components of the AB oscillations are negligible, we derive an expression for the phaseshift and show that the phaseshift is independent of the coherent indirect coupling and monotonically increasing function with respect to the source-drain bias voltage under low source-drain bias voltage conditions. To observe AB oscillations with a large amplitude, we need strong coupling between the QD and the voltage probe. However, this induces the dephasing of the electronic states in the DQD. The competition between such antithetic concepts generates a peak structure when the coupling between the quantum dots and the voltage probe changes.

The outline of this paper is as follows. In Sec. \ref{model}, we introduce a microscopic model Hamiltonian with the three reservoirs ($S$, $D$, and $\varphi$ as shown in Fig. \ref{system}) and the notion of coherent indirect coupling\cite{kubo,tokura}. In Sec. \ref{formulation}, we provide a theoretical formulation based on the nonequilibrium Green's function method\cite{schwinger,keldysh}. In particular, we impose a condition for the voltage probe $\varphi$. Section \ref{result} is devoted to theoretical results for nonlinear transport properties. In Sec. \ref{result2}, we discuss the interplay between the coherent effect of the coupling $\Gamma_{\varphi}$ and the dephasing in relation to the B\"{u}ttiker probe, where $\Gamma_{\varphi}$ is the coupling strength between the QD and the voltage probe. Section \ref{summary} summarizes our results. In Appendix \ref{probability-condition}, we derive the relation that the transmission probability has to satisfy. We show that the linear conductance satisfies the phase symmetry relation in Appendix \ref{derivation-phase}. In Appendix \ref{anti-ap}, we show the antisymmetricity of the lowest-order nonlinear conductance coefficient when the system has mirror symmetry. In Appendix \ref{ap1}, we provide the detailed derivation of the visibility of the AB oscillations in the linear conductance in the limit of $\alpha=1$ and $\Gamma_{\varphi}\to\infty$.

\section{Model\label{model}}
\begin{figure}
\includegraphics[scale=0.6]{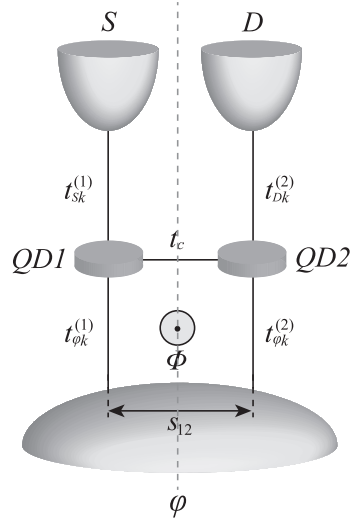}
\caption{\label{system} Schematic diagram of an AB interferometer containing a DQD coupled to three reservoirs. The two QDs couple to three reservoirs, namely the source ($S$), drain ($D$), and voltage probe ($\varphi$). $s_{12}$ is the propagation length of electrons in the voltage probe. $\Phi$ is a magnetic flux threading through the AB interferometer. The dashed line indicates the axis of the mirror symmetry.}
\end{figure}

We consider an AB interferometer containing a DQD coupled to three reservoirs as shown in Fig. \ref{system}. To focus on the coherent charge transport, we neglect the spin degree of freedom. Moreover, we assume that the level spacing is much larger than the source-drain bias voltage, and consider only a single energy level in each QD. The Hamiltonian represents the sum of the following terms: $H=H_R+H_{DQD}+H_T$. The Hamiltonian of the Fermi liquid reservoirs is
\begin{eqnarray}
H_R=\sum_{\nu\in\{S,D,\varphi\}}\sum_k\epsilon_{\nu k}{a_{\nu k}}^{\dagger}a_{\nu k},
\end{eqnarray}
where $\epsilon_{\nu k}$ is the electron energy with wave number $k$ in the reservoir $\nu$, and the operator $a_{\nu k}$ (${a_{\nu k}}^{\dagger}$) annihilates (creates) an electron in the reservoir $\nu$. $H_{DQD}$ describes the isolated DQD,
\begin{eqnarray}
H_{DQD}=\sum_{j=1}^2\epsilon_j{c_j}^{\dagger}c_j+t_c\left({c_1}^{\dagger}c_2+\mbox{h.c.} \right).
\end{eqnarray}
Here $\epsilon_j$ is the energy level of the $j$th QD, and $t_c$ is the direct inter-dot tunnel coupling. The tunneling Hamiltonian between the QDs and the reservoirs is given by
\begin{eqnarray}
H_T&=&\sum_k\left[t_{Sk}^{(1)}{a_{Sk}}^{\dagger}c_1+t_{Dk}^{(2)}{a_{Dk}}^{\dagger}c_2+t_{\varphi k}^{(1)}e^{i\phi/2}{a_{\varphi k}}^{\dagger}c_1+t_{\varphi k}^{(2)}e^{-i\phi/2}{a_{\varphi k}}^{\dagger}c_2+\mbox{h.c.} \right]\nonumber\\
&\equiv&\sum_k\left[t_{Sk}^{(1)}{a_{Sk}}^{\dagger}c_1+t_{Dk}^{(2)}{a_{Dk}}^{\dagger}c_2+\mbox{h.c.} \right]+\sum_k\sum_{j=1}^2\left[t_{\varphi k}^{(j)}(\phi){a_{\varphi k}}^{\dagger}c_j+\mbox{h.c.} \right],
\end{eqnarray}
where $t_{\nu k}^{(j)}$ is the tunneling amplitude between the $j$th QD and the reservoir $\nu$. As an effect of the magnetic flux, we introduced the Peierls phase factors $e^{\pm i\phi/2}$ ($\phi=2\pi\Phi/\Phi_0$ is an AB phase, where $\Phi$ is the magnetic flux threading through an AB ring consisting of the two QDs and the voltage probe as shown in Fig. \ref{system}, and $\Phi_0=h/e$ is the magnetic flux quantum.)

For the reservoir $\nu\in S,D$, the linewidth functions are
\begin{eqnarray}
\Gamma_{jj}^{\nu}(\epsilon)&=&\frac{2\pi}{\hbar}\sum_k|{t_{\nu k}^{(j)}}|^2\delta(\epsilon-\epsilon_{\nu k})\nonumber\\
&\approx &\frac{2\pi}{\hbar}|{t_{\nu}^{(j)}}|^2\rho_{\nu}\nonumber\\
&\equiv&\Gamma_{jj}^{\nu},
\end{eqnarray}
which is assumed to be independent of the energy in the range of interest ($\rho_{\nu}$ is the density of states in the reservoir $\nu$), and when $\nu\in S(D)$, we have $j=1 (2)$. Here we introduced the notation $M_{ij}$, which denotes the $(i,j)$ matrix element of a $2\times2$ matrix $\bm{M}$, where the boldface notation indicates a $2\times2$ matrix whose basis is a localized state in each QD. In contrast, for the reservoir $\varphi$, the linewidth function matrix is not diagonal as follows
\begin{eqnarray}
\Gamma_{ij}^{\varphi}(\epsilon,\phi)&=&\frac{2\pi}{\hbar}\sum_k{t_{\varphi k}^{(i)}}^*(\phi)t_{\varphi k}^{(j)}(\phi)\delta(\epsilon-\epsilon_{\varphi k})\nonumber\\
&\equiv&\Gamma_{ij}^{\varphi}(\phi),
\end{eqnarray}
where we assumed a wide-band limit, namely we neglected the energy dependence. Using the matrix representation, we have
\begin{eqnarray}
\bm{\Gamma}^S=\left(
  \begin{array}{cc}
    \Gamma_S   &  0  \\
    0   &  0  \\
  \end{array}
\right)\ ,\ \bm{\Gamma}^D=\left(
  \begin{array}{cc}
   0    &  0  \\
   0    &  \Gamma_D  \\
  \end{array}
\right)\ ,\ \bm{\Gamma}^{\varphi}(\phi)=\left(
  \begin{array}{cc}
    \Gamma_{\varphi 1}   & \alpha\sqrt{\Gamma_{\varphi 1}\Gamma_{\varphi 2}} e^{-i\phi}   \\
    \alpha \sqrt{\Gamma_{\varphi 1}\Gamma_{\varphi 2}}e^{i\phi}   &  \Gamma_{\varphi 2}  \\
  \end{array}
\right),\label{lwf}
\end{eqnarray}
and the total linewidth function matrix is defined as $\bm{\Gamma}(\phi)=\bm{\Gamma}^S+\bm{\Gamma}^D+\bm{\Gamma}^{\varphi}(\phi)$. When we calculate the linewidth functions from the definitions (4) and (5), we estimate the tunneling amplitude $t_{\nu k}^{(j)}$ in the tunneling Hamiltonian (3) using the Bardeen's formula \cite{bardeen}. In the Bardeen's theory, the tunneling amplitude can be expressed by the wave functions of evanescent mode of the reservoir $\nu$ and a localized electron in the $j$th QD. Then, the tunneling amplitude $t_{\nu k}^{(j)}$ depends on the position of the quantum dot. As a result, the coherent indirect coupling parameter $\alpha$ is a function of the distance between two QDs. This coherent indirect coupling parameter $\alpha$ characterizes the strength of the indirect coupling between two QDs via the voltage probe \cite{kubo,tokura}. The coherent indirect coupling parameter becomes small and changes its sign with increasing distance ($s_{12}$ in Fig. \ref{system}) between the two QDs\cite{kubo}. The importance of the sign of the coherent indirect coupling parameter was pointed out by S. A. Gurvitz\cite{gurvitz}. The influence of the AB effects on the sign of the coherent indirect coupling parameters has been examined experimentally\cite{abe7}. From Eq. (\ref{lwf}), all physical quantities are invariant under the transformation that we change the sign of $\alpha$ and shift the AB phase by $\pi$.

\section{Formulation\label{formulation}}
The tunneling current from the reservoir $S$ to the DQD is given by \cite{current}
\begin{eqnarray}
I_{DQD}(\phi)&=&\frac{e}{h}\int d\epsilon[f_S(\epsilon)-f_D(\epsilon)]\mbox{Tr}\left\{\bm{G}^r(\epsilon,\phi)\bm{\Gamma}^S\bm{G}^a(\epsilon,\phi)\bm{\Gamma}^D \right\}\nonumber\\
&&+\frac{e}{h}\int d\epsilon[f_S(\epsilon)-f_{\varphi}(\epsilon)]\mbox{Tr}\left\{\bm{G}^r(\epsilon,\phi)\bm{\Gamma}^S\bm{G}^a(\epsilon,\phi)\bm{\Gamma}^{\varphi}(\phi) \right\}\nonumber\\
&\equiv&\frac{e}{h}\int d\epsilon[f_S(\epsilon)-f_D(\epsilon)]T_{DS}(\epsilon,\phi)+\frac{e}{h}\int d\epsilon[f_S(\epsilon)-f_{\varphi}(\epsilon)]T_{\varphi S}(\epsilon,\phi),\label{DQD-current}
\end{eqnarray}
where the retarded Green's function is the Fourier transform of
\begin{eqnarray}
G_{ij}^r(t,t')&=&-i\theta(t-t')\langle \{c_i(t),{c_j}^{\dagger}(t') \} \rangle,
\end{eqnarray}
and the advanced Green's function is obtained from the retarded Green's function: $\bm{G}^a(\epsilon,\phi)=[\bm{G}^r(\epsilon,\phi)]^{\dagger}$. $f_{\nu}(\epsilon)$ is the Fermi-Dirac distribution function of the reservoir $\nu$ defined as
\begin{eqnarray}
f_{\nu}(\epsilon)=\frac{1}{e^{(\epsilon-\mu_{\nu})/k_BT}+1},
\end{eqnarray}
where $\mu_{\nu}$ is the electrochemical potential of the reservoir $\nu$, and $T$ is the temperature. In the following discussions, we assume that the source and drain reservoirs have electrochemical potentials $\mu_S=\mu+eV_{SD}/2$ and $\mu_D=\mu-eV_{SD}/2$ with the source-drain bias voltage $V_{SD}$, and $\mu=0$. Here we define the transmission probability from the reservoir $\nu$ to the reservoir $\xi$ as
\begin{eqnarray}
T_{\xi\nu}(\epsilon,\phi)\equiv\mbox{Tr}\left\{\bm{G}^r(\epsilon,\phi)\bm{\Gamma}^{\nu}(\phi)\bm{G}^a(\epsilon,\phi)\bm{\Gamma}^{\xi}(\phi) \right\}.\label{tra-def}
\end{eqnarray}
The electrochemical potential $\mu_{\varphi}$ is determined by the condition that the net current $I_{\varphi}(\phi)$ flowing through the voltage probe $\varphi$ vanishes. Then, we impose following condition to determine $\mu_{\varphi}$
\begin{eqnarray}
I_{\varphi}(\phi)&=&\frac{e}{h}\int d\epsilon[f_{\varphi}(\epsilon,\phi,V_{SD})-f_S(\epsilon)]T_{S\varphi}(\epsilon,\phi)+\frac{e}{h}\int d\epsilon[f_{\varphi}(\epsilon,\phi,V_{SD})-f_D(\epsilon)]T_{D\varphi}(\epsilon,\phi)\nonumber\\
&=&0.\label{floating-condition}
\end{eqnarray}
The reservoir $\varphi$ that satisfies such a condition is equivalent to the B\"{u}ttiker dephasing reservoir \cite{buttiker}.

To calculate the above physical quantities, we need the retarded Green's function. In our model, the retarded Green's function is given by
\begin{eqnarray}
\bm{G}^r(\epsilon,\phi)&=&\left(
  \begin{array}{cc}
     \frac{\epsilon-\epsilon_1}{\hbar}+\frac{i}{2}\Gamma_{11}  &  -\frac{t_c}{\hbar}+\frac{i}{2}\Gamma_{12}(\phi)  \\
     -\frac{t_c}{\hbar}+\frac{i}{2}\Gamma_{21}(\phi)  & \frac{\epsilon-\epsilon_2}{\hbar}+\frac{i}{2}\Gamma_{22}   \\
  \end{array}
\right)^{-1}\nonumber\\
&=&\frac{1}{\Delta(\epsilon,\phi)}\left(
  \begin{array}{cc}
     \frac{\epsilon-\epsilon_2}{\hbar}+\frac{i}{2}\Gamma_{22}  &  \frac{t_c}{\hbar}-\frac{i}{2}\Gamma_{12}(\phi)  \\
     \frac{t_c}{\hbar}-\frac{i}{2}\Gamma_{21}(\phi)  & \frac{\epsilon-\epsilon_1}{\hbar}+\frac{i}{2}\Gamma_{11}   \\
  \end{array}
\right),
\end{eqnarray}
where
\begin{eqnarray}
\Delta(\epsilon,\phi)=\left(\frac{\epsilon-\epsilon_1}{\hbar}+\frac{i}{2}\Gamma_{11} \right)\left(\frac{\epsilon-\epsilon_2}{\hbar}+\frac{i}{2}\Gamma_{22} \right)-\left[\frac{t_c}{\hbar}-\frac{i}{2}\Gamma_{12}(\phi) \right]\left[\frac{t_c}{\hbar}-\frac{i}{2}\Gamma_{21}(\phi) \right].
\end{eqnarray}

In the following, we calculate the linear and the lowest-order nonlinear conductance coefficient with respect to the source-drain bias voltage. In general, the current is expressed as a polynomial function of the source-drain bias voltage $V_{SD}$,
\begin{eqnarray}
I_{DQD}(\phi)=G_{DQD}^{(1)}(\phi)V_{SD}+\frac{1}{2!}G_{DQD}^{(2)}(\phi){V_{SD}}^2+\cdots.\label{current-expansion}
\end{eqnarray}
To calculate the linear and nonlinear conductance coefficient, we focus on the zero temperature condition and employ the following linear approximation for the transmission probability
\begin{eqnarray}
T_{\xi\nu}(\epsilon,\phi)\simeq T_{\xi\nu}(\epsilon=0,\phi)+\left.\frac{\partial T_{\xi\nu}(\epsilon,\phi)}{\partial\epsilon}\right|_{\epsilon=0}\epsilon\label{linear-approximation}
\end{eqnarray}
From the condition (\ref{floating-condition}), the electrochemical potential of the reservoir $\varphi$ is given by
\begin{eqnarray}
\mu_{\varphi}(\phi)&\simeq&\mu_{\varphi}^{(0)}(\phi)+\mu_{\varphi}^{(1)}(\phi)eV_{SD}+\frac{1}{2!}\mu_{\varphi}^{(2)}(\phi)(eV_{SD})^2,\label{chemi}
\end{eqnarray}
where
\begin{eqnarray}
\mu_{\varphi}^{(0)}(\phi)&=&0,\\
\mu_{\varphi}^{(1)}(\phi)&=&\frac{1}{2}\frac{T_{S\varphi}(\epsilon=0,\phi)-T_{D\varphi}(\epsilon=0,\phi)}{T_{S\varphi}(\epsilon=0,\phi)+T_{D\varphi}(\epsilon=0,\phi)},\label{chem1}\\
\mu_{\varphi}^{(2)}(\phi)&=&\frac{1}{4}\frac{\left[\left. \frac{\partial T_{S\varphi}(\epsilon,\phi)}{\partial\epsilon}\right|_{\epsilon=0}+\left. \frac{\partial T_{D\varphi}(\epsilon,\phi)}{\partial\epsilon}\right|_{\epsilon=0} \right]\left[1-\left\{\frac{T_{S\varphi}(\epsilon=0,\phi)-T_{D\varphi}(\epsilon=0,\phi)}{T_{S\varphi}(\epsilon=0,\phi)+T_{D\varphi}(\epsilon=0,\phi)} \right\}^2 \right]}{T_{S\varphi}(\epsilon=0,\phi)+T_{D\varphi}(\epsilon=0,\phi)}\label{chem2}.
\end{eqnarray}
Using Eqs. (\ref{DQD-current}), (\ref{current-expansion}), (\ref{linear-approximation}), and (\ref{chemi}), the linear conductance is given by
\begin{eqnarray}
G_{DQD}^{(1)}(\phi)=\frac{e^2}{h}\left[T_{DS}(\epsilon=0,\phi)+\frac{T_{\varphi S}(\epsilon=0,\phi)T_{D\varphi}(\epsilon=0,\phi)}{T_{S\varphi}(\epsilon=0,\phi)+T_{D\varphi}(\epsilon=0,\phi)} \right],\label{linear-conductance}
\end{eqnarray}
where
\begin{eqnarray}
T_{DS}(\epsilon,\phi)&=&\frac{\hbar\Gamma_S\hbar\Gamma_D}{|\hbar^2\Delta(\epsilon,\phi)|^2}\left[{t_c}^2+\left(\frac{\alpha\hbar\sqrt{\Gamma_{\varphi1}\Gamma_{\varphi2}}}{2} \right)^2+t_c\alpha\hbar\sqrt{\Gamma_{\varphi1}\Gamma_{\varphi2}} \right],\\
T_{\varphi S}(\epsilon,\phi)&=&\frac{1}{|\hbar^2\Delta(\epsilon,\phi)|^2}\left[\hbar\Gamma_S\hbar\Gamma_{\varphi1}\left\{(\epsilon-\epsilon_2)^2+\left(\frac{\hbar\Gamma_D+\hbar\Gamma_{\varphi2}}{2} \right)^2 \right\}\right.\nonumber\\
&&+\hbar\Gamma_S\hbar\Gamma_{\varphi2}\left\{{t_c}^2+\left(\frac{\alpha\hbar\sqrt{\Gamma_{\varphi1}\Gamma_{\varphi2}}}{2} \right)^2+t_c\alpha\hbar\sqrt{\Gamma_{\varphi1}\Gamma_{\varphi2}}\sin\phi \right\}\nonumber\\
&&\left.+2\hbar\Gamma_S\alpha\hbar\sqrt{\Gamma_{\varphi1}\Gamma_{\varphi2}}\left\{(\epsilon-\epsilon_2)t_c\cos\phi-\frac{1}{2}(\hbar\Gamma_D+\hbar\Gamma_{\varphi2})\left(t_c\sin\phi+\frac{\alpha\hbar\sqrt{\Gamma_{\varphi1}\Gamma_{\varphi2}}}{2} \right) \right\} \right],\\
T_{D\varphi}(\epsilon,\phi)&=&\frac{1}{|\hbar^2\Delta(\epsilon,\phi)|^2}\left[\hbar\Gamma_D\hbar\Gamma_{\varphi1}\left\{{t_c}^2+\left(\frac{\alpha\hbar\sqrt{\Gamma_{\varphi1}\Gamma_{\varphi2}}}{2} \right)^2+t_c\alpha\hbar\sqrt{\Gamma_{\varphi1}\Gamma_{\varphi2}}\sin\phi \right\} \right.\nonumber\\
&&+\hbar\Gamma_D\hbar\Gamma_{\varphi2}\left\{(\epsilon-\epsilon_1)^2+\left(\frac{\hbar\Gamma_S+\hbar\Gamma_{\varphi1}}{2} \right)^2 \right\}\nonumber\\
&&\left.+2\hbar\Gamma_D\alpha\hbar\sqrt{\Gamma_{\varphi1}\Gamma_{\varphi2}}\left\{(\epsilon-\epsilon_1)t_c\cos\phi-\frac{1}{2}(\hbar\Gamma_S+\hbar\Gamma_{\varphi1})\left(t_c\sin\phi+\frac{\alpha\hbar\sqrt{\Gamma_{\varphi1}\Gamma_{\varphi2}}}{2} \right) \right\} \right],
\end{eqnarray}
and
\begin{eqnarray}
|\hbar^2\Delta(\epsilon,\phi)|^2&=&\left[(\epsilon-\epsilon_1)(\epsilon-\epsilon_2)-{t_c}^2-\frac{1}{4}\left\{(\hbar\Gamma_S+\hbar\Gamma_{\varphi1})(\hbar\Gamma_D+\hbar\Gamma_{\varphi2})-(\alpha\hbar\sqrt{\Gamma_{\varphi1}\Gamma_{\varphi2}})^2 \right\} \right]^2\nonumber\\
&&+\frac{1}{4}\left[(\epsilon-\epsilon_1)(\hbar\Gamma_D+\hbar\Gamma_{\varphi2})+(\epsilon-\epsilon_2)(\hbar\Gamma_S+\hbar\Gamma_{\varphi1})+2t_c\alpha\hbar\sqrt{\Gamma_{\varphi1}\Gamma_{\varphi2}}\cos\phi \right]^2.
\end{eqnarray}
We confirmed the relation
\begin{eqnarray}
T_{\xi\nu}(\epsilon,-\phi)=T_{\nu\xi}(\epsilon,\phi).\label{trs}
\end{eqnarray}
This is based on time-reversal symmetry. In the discussions in the later sections, we sometimes consider a highly symmetric situation, namely $\epsilon_1=\epsilon_2\equiv\epsilon_d$, $\Gamma_S=\Gamma_D\equiv\Gamma$, and $\Gamma_{\varphi1}=\Gamma_{\varphi2}\equiv\Gamma_{\varphi}$ (mirror symmetry with respect to the dashed line in Fig. \ref{system}). Under this condition, we can have following additional relation:
\begin{eqnarray}
T_{\varphi S}(\epsilon,\phi)=T_{D\varphi}(\epsilon,\phi).\label{mirror}
\end{eqnarray}

Before we discuss nonlinear transport in the next section, we consider linear transport. From the conservation of probability, we have the following relation
\begin{eqnarray}
\sum_{\xi\neq\nu}T_{\nu\xi}(\epsilon,\phi)=\sum_{\xi\neq\nu}T_{\xi\nu}(\epsilon,\phi).\label{aps1}
\end{eqnarray}
The derivation of this relation is given in Appendix \ref{probability-condition}. As proven in Appendix \ref{derivation-phase}, using this relation for arbitrary parameters, we find that the linear conductance satisfies the Onsager-Casimir symmetry relation \cite{onsager,casimir}
\begin{eqnarray}
G_{DQD}^{(1)}(-\phi)=G_{DQD}^{(1)}(\phi).\label{phase-symmetry}
\end{eqnarray}
This satisfies the B\"{u}ttiker's result for the linear conductance\cite{multi3}.

\begin{figure}
\includegraphics[scale=0.6]{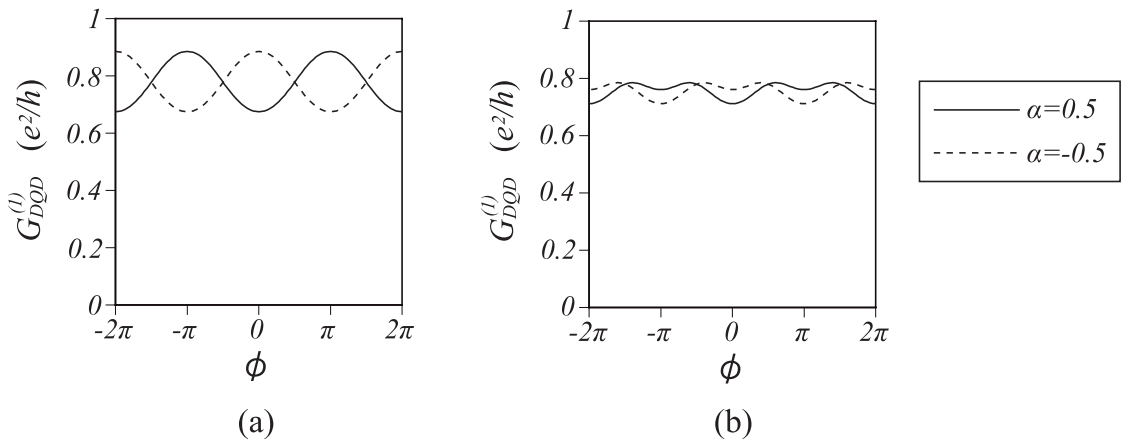}
\caption{\label{linear} AB oscillations of the linear conductance under the condition of mirror symmetry for $\Gamma=\Gamma_{\varphi}$, $t_c/\hbar\Gamma=-1$. (a) $\epsilon_d/\hbar\Gamma=1$. (b) $\epsilon_d/\hbar\Gamma=0.2$. The solid and broken lines indicate cases where $\alpha=0.5$ and $\alpha=-0.5$, respectively.}
\end{figure}

In Fig. \ref{linear}, we show the numerical results for the AB oscillations of the linear conductance under the condition of mirror symmetry. In Fig. \ref{linear} (a), we plot the AB oscillations of the linear conductance when $\Gamma=\Gamma_{\varphi}$, $t_c/\hbar\Gamma=-1$, and $\epsilon_d/\hbar\Gamma=1$. In this case, the convex shape of the linear conductance at $\phi=0$ depends on the sign of the coherent indirect coupling parameter $\alpha$. Similarly, we show the AB oscillations of the linear conductance when $\Gamma=\Gamma_{\varphi}$, $t_c/\hbar\Gamma=-1$, and $\epsilon_d/\hbar\Gamma=0.2$. Under this condition, we find that the convex shape of the linear conductance at $\phi=0$ is independent of the sign of the coherent indirect coupling parameter $\alpha$.

\section{Phase symmetry breaking\label{result}}
In this section, we discuss the phase symmetry breaking in the nonlinear transport regime. We derive the condition under which the phase symmetry is broken by calculating the lowest-order nonlinear conductance coefficient.

\subsection{Nonlinear transport in weak inter-dot coherent coupling}
Here we discuss the nonlinear transport under a finite source-drain bias voltage. Before discussing the general properties of the lowest-order nonlinear conductance coefficient, we focus on the weak inter-dot coupling situation where $|\alpha|\hbar\Gamma_{\varphi},|t_c|\ll\hbar\Gamma$ to remove the higher harmonic components of the AB oscillations and obtain an intuitive picture of the phaseshift. Under the mirror symmetry condition, the tunneling current through a DQD is written as
\begin{eqnarray}
I_{DQD}(\phi)&\simeq&I_0-\alpha t_cI_1\cos\phi-\alpha t_cI_2\sin\phi\\
&=&I_0-\alpha t_c\sqrt{{I_1}^2+{I_2}^2}\cos\left(\phi-\Delta\phi \right),
\end{eqnarray}
where
\begin{eqnarray}
I_0&=&\frac{e^2}{h}\frac{2\hbar\Gamma\hbar\Gamma_{\varphi}}{4{\epsilon_d}^2+(\hbar\Gamma+\hbar\Gamma_{\varphi})^2}V_{SD},\\
I_1&=&\frac{e^2}{h}\frac{16\epsilon_d\hbar\Gamma\hbar\Gamma_{\varphi}\Lambda}{\left[4{\epsilon_d}^2+(\hbar\Gamma+\hbar\Gamma_{\varphi})^2 \right]^3}V_{SD},\\
I_2&=&\frac{e^3}{h}\frac{32\epsilon_d(\hbar\Gamma)^2\hbar\Gamma_{\varphi}}{4{\epsilon_d}^2+(\hbar\Gamma+\hbar\Gamma_{\varphi})^2}{V_{SD}}^2,
\end{eqnarray}
and
\begin{eqnarray}
\Lambda\equiv 4{\epsilon_d}^2+(\hbar\Gamma-3\hbar\Gamma_{\varphi})(\hbar\Gamma+\hbar\Gamma_{\varphi}).
\end{eqnarray}
Here $\Delta\phi$ ($-\pi\le \Delta\phi\le\pi$) satisfies the relations
\begin{eqnarray}
\cos(\Delta\phi)=\frac{I_1}{\sqrt{{I_1}^2+{I_2}^2}}\ ,\ \sin(\Delta\phi)=\frac{I_2}{\sqrt{{I_1}^2+{I_2}^2}},\label{cos-sin}
\end{eqnarray}
Thus, we can directly derive the expression of the phaseshift from Eq. (\ref{cos-sin})
\begin{eqnarray}
\Delta\phi&=&\tan^{-1}\left(\frac{I_2}{I_1} \right)\\
&=&\tan^{-1}\left(\frac{2\hbar\Gamma}{\Lambda}eV_{SD} \right).\label{phaseshift}
\end{eqnarray}
When $V_{SD}\neq 0$, the phaseshift is finite and the phase symmetry is broken. From Eq. (\ref{phaseshift}) we find that the phaseshift is independent of the inter-dot couplings, namely $t_c$ and $\alpha\hbar\Gamma_{\varphi}$, and a monotonic function with respect to the source-drain bias voltage $V_{SD}$. In this expression, the phaseshift is defined in the range of $-\frac{\pi}{2}\le\Delta\phi\le\frac{\pi}{2}$. However, the original $\Delta\phi$ is $-\pi\le\Delta\phi\le\pi$ range. When $I_1$ is negative, namely $\frac{\pi}{2}<\Delta\phi\le\pi$ or $-\pi<\Delta\phi<-\frac{\pi}{2}$, we can express the results in the $-\frac{\pi}{2}\le\Delta\phi\le\frac{\pi}{2}$ range by shifting the phaseshift by $\pi$ and changing the sign of the current.

\begin{figure}[htbp]
  \begin{center}
    \includegraphics[scale=0.6]{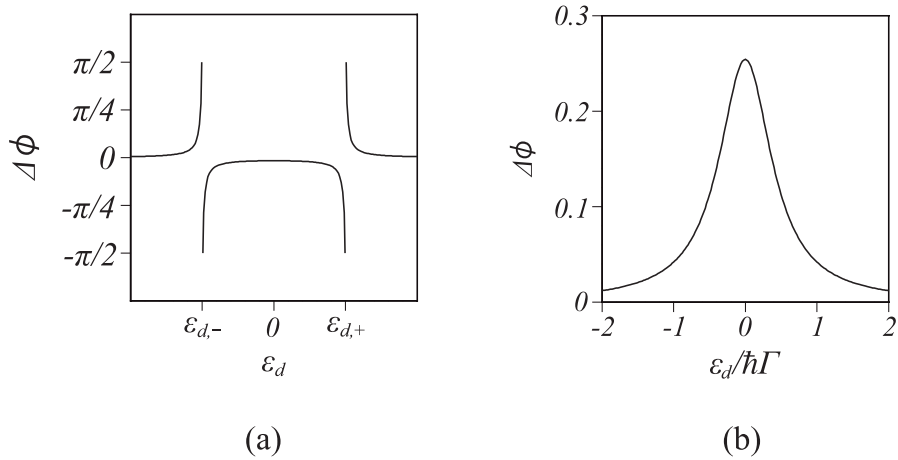}
  \end{center}
  \caption{QD energy $\epsilon_d$ dependence of the phaseshift $\Delta\phi$ for $eV_{SD}/\hbar\Gamma=0.1$. (a) Phaseshift for $\Gamma_{\varphi}=\Gamma$ ($\hbar\Gamma<3\hbar\Gamma_{\varphi}$). (b) $\Gamma_{\varphi}=0.1\Gamma$ ($\hbar\Gamma\ge3\hbar\Gamma_{\varphi}$).}
  \label{fig:phaseshift-weak-coupling.eps}
\end{figure}

\begin{figure}[htbp]
  \begin{center}
    \includegraphics[scale=0.5]{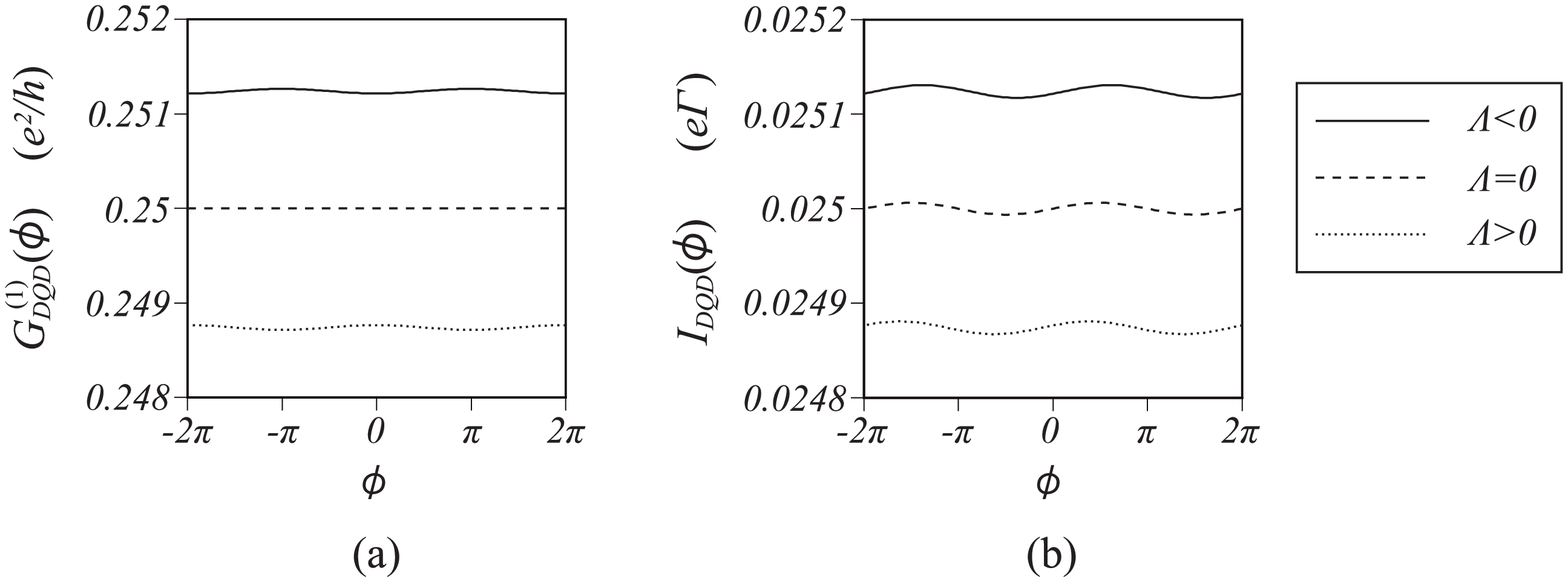}
  \end{center}
  \caption{AB oscillations of the linear conductance and the tunneling current through a DQD where $\Lambda>0$ ($\Gamma_{\varphi}=0.99\Gamma$), $\Lambda=0$ ($\Gamma_{\varphi}=\Gamma$), and $\Lambda<0$ ($\Gamma_{\varphi}=1.01\Gamma$) when $\epsilon_d/\hbar\Gamma=1$, $t_c/\hbar\Gamma=-0.1$, $\alpha=0.1$, and $eV_{SD}/\hbar\Gamma=0.1$. (a) Linear conductance. (b) Tunneling current.}
  \label{fig:LC-IDQD-AB.eps}
\end{figure}

Here we consider the QD energy dependence of the phaseshift. First we consider the case where $\hbar\Gamma<3\hbar\Gamma_{\varphi}$. We have
\begin{eqnarray*}
\lim_{\epsilon_d\to|\epsilon_{d,\pm}|\pm 0}\Delta\phi=\pm\frac{\pi}{2},
\end{eqnarray*}
and thus we find that the phaseshift jumps by $\pi$ at
\begin{eqnarray*}
\epsilon_{d,\pm}=\pm\frac{\sqrt{(3\hbar\Gamma_{\varphi}-\hbar\Gamma)(\hbar\Gamma+\hbar\Gamma_{\varphi})}}{2}
\end{eqnarray*}
as shown in Fig. \ref{fig:phaseshift-weak-coupling.eps} (a) when $\Gamma_{\varphi}=\Gamma$ and $eV_{SD}/\hbar\Gamma=0.1$. For $\Lambda=0$, the phaseshift is not defined since the linear conductance does not exhibit an Aharonov-Bohm (AB) oscillation as shown in Fig. \ref{fig:LC-IDQD-AB.eps}(a) dashed-line. Although the phaseshift jumps by $\pi$ at $\Lambda=0$, the tunneling current changes smoothly as shown in Fig. \ref{fig:LC-IDQD-AB.eps} (b). This phaseshift jump is based on the fact that the convex shape of the linear conductance changes at $\Lambda=0$ (see Fig. \ref{fig:LC-IDQD-AB.eps} (a)).

Next we consider the case where $\hbar\Gamma\ge 3\hbar\Gamma_{\varphi}$. We find that the phaseshift is always positive, a smooth function with respect to the QD energy $\epsilon_d$, and has its maximum value at $\epsilon_d=0$. Fig. \ref{fig:phaseshift-weak-coupling.eps} (b) shows the numerical result when $\Gamma_{\varphi}=0.1\Gamma$ and $eV_{SD}/\hbar\Gamma=0.1$.

\subsection{Electrochemical potential of voltage probe in weak inter-dot coherent coupling}
Here we discuss the electrochemical potential of the voltage probe $\varphi$, which can be measured experimentally. Under the mirror symmetry condition, using Eqs. (\ref{chem1}), (\ref{chem2}), (\ref{trs}), and (\ref{mirror}) we can show that
\begin{eqnarray}
\mu_{\varphi}^{(1)}(-\phi)=-\mu_{\varphi}^{(1)}(\phi)\ ,\ \mu_{\varphi}^{(2)}(-\phi)=\mu_{\varphi}^{(2)}(\phi).
\end{eqnarray}
Furthermore, with the weak inter-dot coupling regime, Eqs. (\ref{chem1}) and (\ref{chem2}) are
\begin{eqnarray}
\mu_{\varphi}^{(1)}(\phi)&=&\frac{2\hbar\Gamma\alpha t_c}{4{\epsilon_d}^2+(\hbar\Gamma+\hbar\Gamma_{\varphi})^2}\sin\phi,\\
\mu_{\varphi}^{(2)}(\phi)&=&\left[\frac{2\epsilon_d}{4{\epsilon_d}^2+(\hbar\Gamma+\hbar\Gamma_{\varphi})^2}-\frac{2\{16{\epsilon_d}^4-48{\epsilon_d}^2\hbar\Gamma_{\varphi}(\hbar\Gamma+\hbar\Gamma_{\varphi})-(\hbar\Gamma-3\hbar\Gamma_{\varphi})(\hbar\Gamma+\hbar\Gamma_{\varphi})^3 \}\alpha t_c}{\{4{\epsilon_d}^2+(\hbar\Gamma+\hbar\Gamma_{\varphi})^2 \}^3}\cos\phi \right].\label{chem22}
\end{eqnarray}
From these results, we find that $\mu_{\varphi}$ vanishes at $\phi=0$ in the linear transport regime and $\mu_{\varphi}^{(2)}$ makes a contribution that is independent of the phase $\phi$ and inter-dot coherent coupling $t_c$ and $\alpha\hbar\Gamma_{\varphi}$. $\mu_{\varphi}^{(1)}$ and $\mu_{\varphi}^{(2)}$ are observables and their AB oscillations can be easily detected in experiments. In particular, it is interesting to note that the sign of the 1st term in Eq. (\ref{chem22}) depends only on the QD energy $\epsilon_d$. Therefore, the sign of the phase-independent contribution of $\mu_{\varphi}^{(2)}(\phi)$ is positive (negative) when the QD energy is above (below) the Fermi level.

\subsection{General properties of nonlinear conductance coefficient}
Here we discuss more general properties of the lowest-order nonlinear conductance coefficient in Eq. (\ref{current-expansion}),
\begin{eqnarray}
G_{DQD}^{(2)}(\phi)&=&\frac{e^3}{h}\frac{1}{4}\left[1-\left\{\frac{T_{S\varphi}(\epsilon=0,\phi)-T_{D\varphi}(\epsilon=0,\phi)}{T_{S\varphi}(\epsilon=0,\phi)+T_{D\varphi}(\epsilon=0,\phi)} \right\}^2 \right]\nonumber\\
&&\times\left[\left.\frac{\partial T_{\varphi S}(\epsilon,\phi)}{\partial\epsilon}\right|_{\epsilon=0}-\frac{\left.\frac{\partial T_{S\varphi}(\epsilon,\phi)}{\partial\epsilon}\right|_{\epsilon=0}+\left.\frac{\partial T_{D\varphi}(\epsilon,\phi)}{\partial\epsilon}\right|_{\epsilon=0}}{T_{S\varphi}(\epsilon=0,\phi)+T_{D\varphi}(\epsilon=0,\phi)}T_{\varphi S}(\epsilon=0,\phi) \right].\label{nlc}
\end{eqnarray}
As shown in Appendix \ref{anti-ap}, under the mirror symmetry condition, we find that the lowest-order nonlinear conductance coefficient is asymmetric with respect to the flux
\begin{eqnarray}
G_{DQD}^{(2)}(-\phi)=-G_{DQD}^{(2)}(\phi). 
\end{eqnarray}
Therefore, the lowest-order nonlinear conductance coefficient has no symmetric component. This contrasts with the lowest-order nonlinear conductance coefficient in two-terminal systems\cite{two-terminal8,nakamura} and Mach-Zehnder interferometers\cite{mz}, which has symmetric and antisymmetric components.

The position of the current peak or dip at zero magnetic field shifts since the phase symmetry is broken. To examine the direction of such a phaseshift, we estimate the 1st-order differential coefficient for the lowest-order nonlinear conductance coefficient at $\phi=0$
\begin{eqnarray}
\left.\frac{\partial G_{DQD}^{(2)}(\phi)}{\partial\phi}\right|_{\phi=0}&=&-\frac{1}{2}\frac{e^3}{h}\frac{T_{\varphi S}(\epsilon=0,\phi=0)-T_{\varphi D}(\epsilon=0,\phi=0)}{T_{\varphi S}(\epsilon=0,\phi=0)+T_{\varphi D}(\epsilon=0,\phi=0)}\frac{\left. \frac{\partial T_{\varphi S}(\epsilon=0,\phi)}{\partial\phi}\right|_{\phi=0}-\left. \frac{\partial T_{\varphi D}(\epsilon=0,\phi)}{\partial\phi}\right|_{\phi=0}}{T_{\varphi S}(\epsilon=0,\phi=0)+T_{\varphi D}(\epsilon=0,\phi=0)}\nonumber\\
&&\times\left[\left. \frac{\partial T_{S\varphi}(\epsilon,\phi=0)}{\partial\epsilon}\right|_{\epsilon=0}-\frac{\left.\frac{\partial T_{\varphi S}(\epsilon,\phi=0)}{\partial\epsilon}\right|_{\epsilon=0}+\left.\frac{\partial T_{\varphi D}(\epsilon,\phi=0)}{\partial\epsilon}\right|_{\epsilon=0}}{T_{\varphi S}(\epsilon=0,\phi=0)+T_{\varphi D}(\epsilon=0,\phi=0)}T_{S\varphi}(\epsilon=0,\phi=0) \right]\nonumber\\
&&+\frac{1}{4}\frac{e^3}{h}\left[1-\left\{\frac{T_{\varphi S}(\epsilon=0,\phi=0)-T_{\varphi D}(\epsilon=0,\phi=0)}{T_{\varphi S}(\epsilon=0,\phi=0)+T_{\varphi D}(\epsilon=0,\phi=0)} \right\}^2 \right]\nonumber\\
&&\times\left[\left.\frac{\partial^2T_{S\varphi}(\epsilon,\phi)}{\partial\phi\partial\epsilon}\right|_{\epsilon=0,\phi=0}-\frac{\left.\frac{\partial T_{S\varphi}(\epsilon=0,\phi)}{\partial\phi}\right|_{\phi=0}\left\{\left. \frac{\partial T_{\varphi S}(\epsilon,\phi)}{\partial\epsilon}\right|_{\epsilon=0,\phi=0}+\left. \frac{\partial T_{\varphi D}(\epsilon,\phi)}{\partial\epsilon}\right|_{\epsilon=0,\phi=0} \right\} }{T_{\varphi S}(\epsilon=0,\phi=0)+T_{\varphi D}(\epsilon=0,\phi=0)} \right].
\end{eqnarray}
In general, we have
\begin{eqnarray}
\left.\frac{\partial G_{DQD}^{(2)}(\phi)}{\partial\phi}\right|_{\phi=0}\neq0.
\end{eqnarray}
For clarity, we consider the mirror symmetry and thus obtain
\begin{eqnarray}
\left. \frac{\partial G_{DQD}^{(2)}(\phi)}{\partial\phi}\right|_{\phi=0}=\frac{e^3}{h}\frac{(\hbar\Gamma)^3(\hbar\Gamma_{\varphi})^2}{|\hbar^2\Delta(\epsilon=0,\phi=0)|^4}t_c\alpha(t_c\alpha-\epsilon_d).\label{direction-condition}
\end{eqnarray}

From Eq. (\ref{direction-condition}), the factor $t_c\alpha(t_c\alpha-\epsilon_d)$ determines the sign of $\left. \frac{\partial G_{DQD}^{(2)}(\phi)}{\partial\phi}\right|_{\phi=0}$. Then, we define this factor as $S(\alpha)$. In a recent experiment, it was reported that the sign of the coherent indirect coupling parameter $\alpha$ can be changed by tuning the gate voltage\cite{abe7}. In the following, as an example, we investigate the direction of the phaseshift when we change the sign of $\alpha$. Then we estimate
\begin{eqnarray}
S(\alpha)S(-\alpha)=(t_c\alpha)^2[(t_c\alpha)^2-{\epsilon_d}^2].
\end{eqnarray}

\begin{figure}
\includegraphics[scale=0.6]{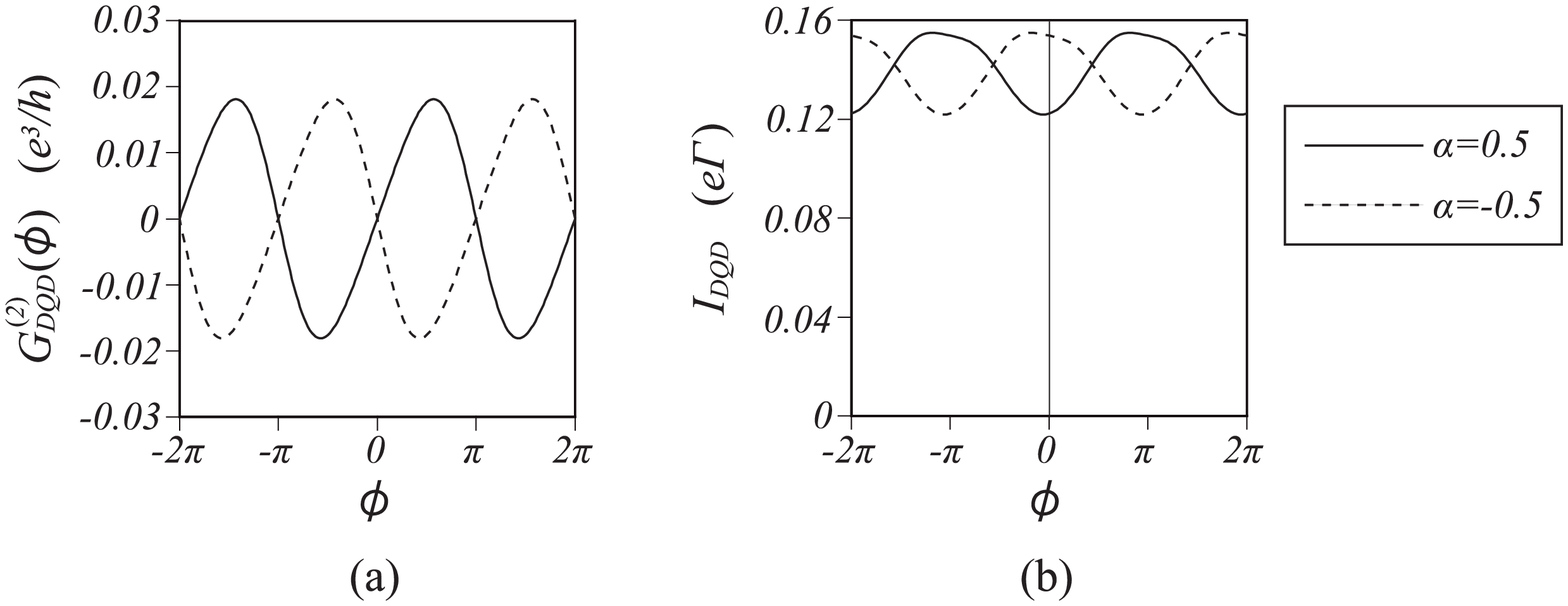}
\caption{\label{depend} AB oscillations of the lowest-order nonlinear conductance coefficient when $\Gamma_S=\Gamma_D=\Gamma_{\varphi}\equiv\Gamma$, $t_c/\hbar\Gamma=-1$, and $\epsilon_d/\hbar\Gamma=1$. The solid and broken lines indicate cases where $\alpha=0.5$ and $\alpha=-0.5$, respectively.}
\end{figure}

\begin{figure}
\includegraphics[scale=0.6]{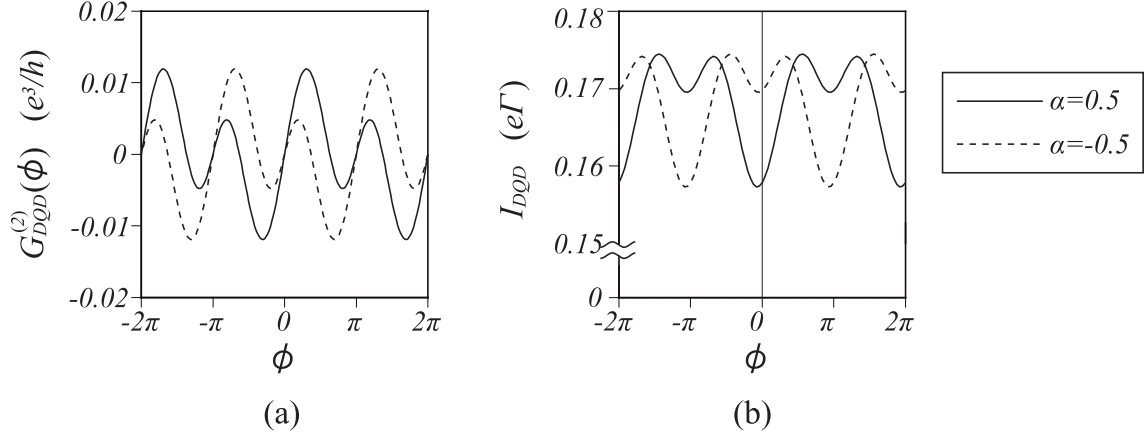}
\caption{\label{independent} AB oscillations of the lowest-order nonlinear conductance coefficient when $\Gamma_S=\Gamma_D=\Gamma_{\varphi}\equiv\Gamma$, $t_c/\hbar\Gamma=-1$, and $\epsilon_d/\hbar\Gamma=0.2$. The solid and broken lines indicate cases where $\alpha=0.5$ and $\alpha=-0.5$, respectively.}
\end{figure}

When $|t_c\alpha|<|\epsilon_d|$, the slope of the lowest-order nonlinear conductance coefficient at $\phi=0$ depends on the sign of $\alpha$. As an example, we show the AB oscillations of the lowest-order nonlinear conductance coefficient in Fig. \ref{depend} (a) when $\Gamma_S=\Gamma_D=\Gamma_{\varphi}$, $t_c/\hbar\Gamma=-1$, $|\alpha|=0.5$, and $\epsilon_d/\hbar\Gamma=1$.
In contrast, when $|t_c\alpha|>|\epsilon_d|$, the slope of the lowest-order nonlinear conductance coefficient at $\phi=0$ is independent of the sign of $\alpha$. As an example, we plot the AB oscillations of the lowest-order nonlinear conductance coefficient in Fig. \ref{independent} (a) when $\Gamma_S=\Gamma_D=\Gamma_{\varphi}\equiv\Gamma$, $t_c/\hbar\Gamma=-1$, $|\alpha|=0.5$, and $\epsilon_d/\hbar\Gamma=0.2$.

From the above results, we consider the direction of the phaseshift for the AB oscillations in the current through a DQD. In Fig. \ref{depend} (b), we plot the AB oscillations in the current through a DQD when $\epsilon_d/\hbar\Gamma=1$, $t_c/\hbar\Gamma=-1$, $\Gamma_S=\Gamma_D=\Gamma_{\varphi}\equiv\Gamma$, and $|\alpha|=0.5$, namely $|t_c\alpha|<|\epsilon_d|$. For $\alpha=0.5$, the conductance exhibits a dip at $\phi=0$ as shown in Fig. \ref{linear} (a). According to Fig. \ref{depend}, the position of this dip shifts to the negative phase direction in a nonlinear transport regime. Similarly, for $\alpha=-0.5$, we have the conductance peak at $\phi=0$ as shown in Fig. \ref{linear} (a). According to Fig. \ref{depend} (a), the position of this peak should shift to the negative phase direction in a nonlinear transport regime. The direction of the phaseshift for the current through a DQD at $eV_{SD}/\hbar\Gamma=2$ is consistent with a prediction for the lowest-order nonlinear conductance coefficient as shown in Fig. \ref{depend} (b). 

Next we consider the situation when $\epsilon_d/\hbar\Gamma=0.2$, $t_c/\hbar\Gamma=-1$, $\Gamma_S=\Gamma_D=\Gamma_{\varphi}\equiv\Gamma$, and $|\alpha|=0.5$, namely $|t_c\alpha|>|\epsilon_d|$. For both $\alpha=0.5$ and $\alpha=-0.5$, the conductance shows a dip at $\phi=0$ as shown in Fig. \ref{linear} (b). According to Fig. \ref{independent} (a), the position of these dips should shift to the negative phase direction in the nonlinear transport regime. The direction of the phaseshift for the current through a DQD at $eV_{SD}/\hbar\Gamma=2$ is consistent with a prediction for the lowest-order nonlinear conductance coefficient as shown in Fig. \ref{independent} (a).

\begin{figure}
\includegraphics[scale=0.6]{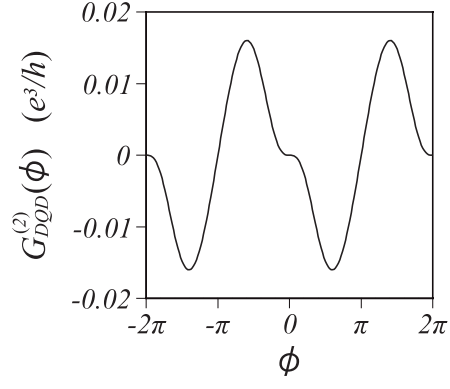}
\caption{\label{inflection-point} Behavior of the lowest-order nonlinear conductance coefficient under the mirror symmetry condition when $t_c\alpha=\epsilon_d$ ($\epsilon_d/\hbar\Gamma=-0.5$, $t_c/\hbar\Gamma=-1$, $\alpha=0.5$, and $\Gamma_{\varphi}=\Gamma$). $\phi=0$ is an inflection point, and the lowest-order nonlinear conductance coefficient has an antisymmetricity under the mirror symmetry condition ($\epsilon_1=\epsilon_2\equiv\epsilon_d$, $\Gamma_S=\Gamma_D\equiv\Gamma$, and $\Gamma_{\varphi1}=\Gamma_{\varphi2}\equiv\Gamma_{\varphi}$).}
\end{figure}

From Eq. (\ref{direction-condition}), when $t_c\alpha=\epsilon_d$, we have $\left.\frac{\partial G_{DQD}^{(2)}(\phi)}{\partial\phi}\right|_{\phi=0}=0$. However, the lowest-order nonlinear conductance coefficient does not have an extreme value at $\phi=0$ since we have $\left.\frac{\partial^2 G_{DQD}^{(2)}(\phi)}{\partial\phi^2}\right|_{\phi=0}=0$ under the same condition. As a result, $\phi=0$ is an inflection point as shown in Fig. \ref{inflection-point} when $\epsilon_d/\hbar\Gamma=-0.5$, $t_c/\hbar\Gamma=-1$, $\alpha=0.5$, and $\Gamma_{\varphi}=\Gamma$.

\section{Dephasing effects of voltage probe $\varphi$\label{result2}}
So far we have discussed the phaseshift in the nonlinear transport regime. In this section, we consider the amplitude of the AB oscillations in the transport properties to study dephasing effects induced by the voltage probe. To observe the AB oscillation, we need the coupling $\Gamma_{\varphi}$. However, this causes dephasing of the electronic states in the DQD. In this section, we study the competition between these two antithetic concepts. In particular, we examine the $\alpha$ and $\Gamma_{\varphi}$ dependences of the AB oscillations in the linear and nonlinear conductances. For simplicity, we focus on the mirror symmetry in this section.

\subsection{Linear conductance}
\begin{figure}
\includegraphics[scale=0.6]{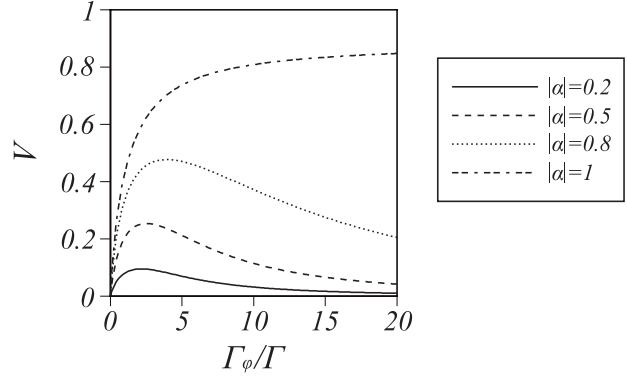}
\caption{\label{visibility} Visibility of the AB oscillations in the linear conductance as a function of $\Gamma_{\varphi}$ for various $|\alpha|$ values when $\epsilon_d/\hbar\Gamma=-t_c/\hbar\Gamma=1$.}
\end{figure}

Here we discuss the AB oscillations in the linear conductance. To investigate the interplay between the two antithetic concepts as mentioned above, we examine how the coherence is modulated as the coupling $\Gamma_{\varphi}$ increases. As a physical quantity that characterizes the coherence, we define the visibility of the AB oscillation as follows
\begin{eqnarray}
V=\frac{G_{DQD,max}^{(1)}-G_{DQD,min}^{(1)}}{G_{DQD,max}^{(1)}+G_{DQD,min}^{(1)}},\label{visibility-definition}
\end{eqnarray}
where $G_{DQD,max}^{(1)}$ and $G_{DQD,min}^{(1)}$ correspond to the maximum and minimum values in the AB oscillations of the linear conductance, respectively. In Fig. \ref{visibility}, we show the coupling $\Gamma_{\varphi}$ dependences of the visibility of the AB oscillations in the linear conductance for various $|\alpha|$ values when $\epsilon_d/\hbar\Gamma=-t_c/\hbar\Gamma=1$.
In the weak coupling regime, the visibility increases monotonically as the coupling $\Gamma_{\varphi}$ increases. This result reveals that we need a stronger coupling $\Gamma_{\varphi}$ to observe the AB oscillations with higher visibility. However, for $|\alpha|\neq 1$, the visibility has a maximum value, and the visibility decreases with $\Gamma_{\varphi}$ in the strong coupling regime. This result means that the strong coupling $\Gamma_{\varphi}$ gives rise to the dephasing, and leads to the loss of the coherence. The interplay between these two features provides the maximum visibility as shown in Fig. \ref{visibility}. Moreover, in the limit of $\Gamma_{\varphi}\gg\Gamma$, the visibility of the AB oscillations has the following leading term in an asymptotic series for $|\alpha|\neq 1$
\begin{eqnarray}
V_{\Gamma_{\varphi}\to\infty}\sim\frac{8|\epsilon_dt_c\alpha|(3+\alpha^2)}{(1-\alpha^2)^2}\left(\frac{1}{\hbar\Gamma_{\varphi}} \right)^2.\label{limit-visibility}
\end{eqnarray}
$V_{\Gamma_{\varphi}\to\infty}$ decreases with $(\Gamma_{\varphi})^{-2}$ with a monotonically increasing coefficient $|\alpha|$.

In contrast, for $|\alpha|=1$, there is no maximum visibility, and the visibility increases monotonically as the coupliing $\Gamma_{\varphi}$ increases. In the limit of infinite $\Gamma_{\varphi}$, the visibility of the AB oscillations with $|\alpha|=1$ can be expressed as
\begin{eqnarray}
V_{\Gamma_{\varphi}\to\infty}=\frac{\frac{e^2}{h}-\min\left(G_{DQD}^{(1)}(\phi=0),G_{DQD}^{(1)}(\phi=\pi) \right)}{\frac{e^2}{h}+\min\left(G_{DQD}^{(1)}(\phi=0),G_{DQD}^{(1)}(\phi=\pi) \right)},\label{visibility-limit}
\end{eqnarray}
where
\begin{eqnarray}
G_{DQD}^{(1)}(\phi=0)&=&\frac{e^2}{h}\frac{(\hbar\Gamma)^2}{4(\epsilon_d-t_c)^2+(\hbar\Gamma)^2},\\
G_{DQD}^{(1)}(\phi=\pi)&=&\frac{e^2}{h}\frac{(\hbar\Gamma)^2}{4(\epsilon_d+t_c)^2+(\hbar\Gamma)^2}.
\end{eqnarray}
In the situation shown in Fig. \ref{visibility}, we have $V_{\Gamma_{\varphi}\to\infty}=8/9$. Moreover, when $\Gamma\neq 0$, using the relation
\begin{eqnarray}
0<\frac{(\hbar\Gamma)^2}{4(\epsilon_d\pm t_c)^2+(\hbar\Gamma)^2}\le 1,
\end{eqnarray}
we can prove that the visibility $V_{\Gamma_{\varphi}\to\infty}<1$ from Eq. (\ref{visibility-limit}). The derivation of the expression of the visibility in Eq. (\ref{visibility-limit}) is given in Appendix \ref{ap1}. Such exceptional behavior at $|\alpha|=1$ can be explained as follows. With the tunnel-coupled symmetric and antisymmetric states as a basis, the linewidth function matrices are given by
\begin{eqnarray}
\bm{\Gamma}^S=\frac{\Gamma}{2}\left(
  \begin{array}{cc}
    1   & 1   \\
    1   & 1   \\
  \end{array}
\right)\ ,\ \bm{\Gamma}^D=\frac{\Gamma}{2}\left(
  \begin{array}{cc}
    1   & -1   \\
    -1   & 1   \\
  \end{array}
\right)\ ,\ \bm{\Gamma}^{\varphi}(\phi)=\Gamma_{\varphi}\left(
  \begin{array}{cc}
    1+\alpha\cos\phi   &  i\alpha\sin\phi  \\
    -i\alpha\sin\phi   &  1-\alpha\cos\phi  \\
  \end{array}
\right).
\end{eqnarray}
Then, the coupling strength between the symmetric (antisymmetric) state and the voltage probe is characterized by $\Gamma_{\varphi}(1+\alpha\cos\phi)$ ($\Gamma_{\varphi}(1-\alpha\cos\phi)$). Thus, at $\phi=0$ ($\phi=\pi$), the antisymmetric (symmetric) state is \textit{dephasing-free} from the voltage probe, and the coupling $\Gamma_{\varphi}$ helps to enhance the coherence. As a result, when $\epsilon_d+t_c=0$ ($\epsilon_d-t_c=0$), the resonant tunneling process is realized through the symmetric (antisymmetric) state in the series-coupled DQD, and the linear conductance has a  value of $e^2/h$. This is the origin of the high visibility in the limit of $|\alpha|=1$ and $\Gamma_{\varphi}\to\infty$. In contrast, for $|\alpha|\neq1$, both the symmetric and antisymmetric states are dephased by the coupling $\Gamma_{\varphi}$ with the voltage probe. Consequently, the linear conductance vanishes without depending on $\phi$, and the visibility of the AB oscillations becomes zero.

\subsection{Lowest-order nonlinear conductance coefficient}
\begin{figure}
\includegraphics[scale=0.4]{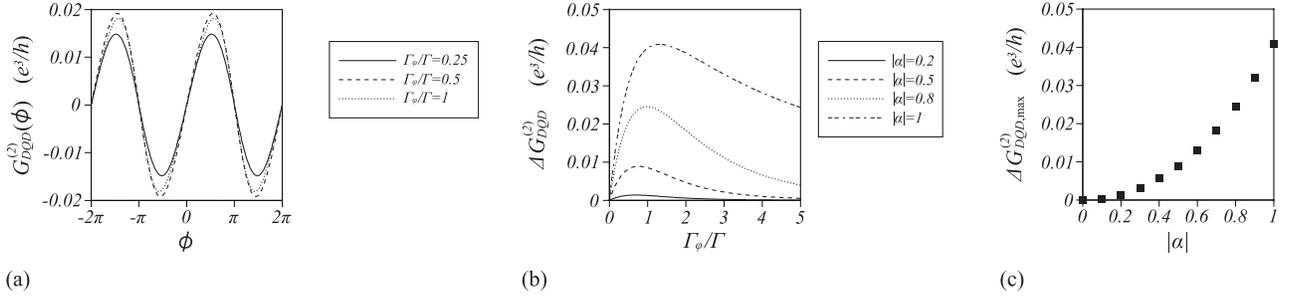}
\caption{\label{NLC-GaP-alpha} AB oscillation of the lowest-order nonlinear conductance coefficient $G_{DQD}^{(2)}(\phi)$ and $\Gamma_{\varphi}$ and the $|\alpha|$ dependences of its amplitude when $\epsilon_d/\hbar\Gamma=1$, $t_c/\hbar\Gamma=-1$. (a) AB oscillations of $G_{DQD}^{(2)}(\phi)$ for $\alpha=0.5$. (b) $\Gamma_{\varphi}$ dependences of $\Delta G_{DQD}^{(2)}$ for various $|\alpha|$ values. (c) $|\alpha|$ dependence of the maximal value $\Delta G_{DQD,{\scriptsize \mbox{max}}}^{(2)}$ in (b).}
\end{figure}

Here we discuss the lowest-order nonlinear conductance coefficient $G_{DQD}^{(2)}(\phi)$ with respect to the source-drain bias voltage. In Fig. \ref{NLC-GaP-alpha} (a), we plot the AB oscillations of $G_{DQD}^{(2)}(\phi)$ for $\epsilon_d/\hbar\Gamma=1$, $t_c/\hbar\Gamma=-1$ and $\alpha=0.5$ for various $\Gamma_{\varphi}$ The amplitude of the AB oscillation depends on $\Gamma_{\varphi}$ values. We discuss the $\Gamma_{\varphi}$ dependences for the amplitude of the AB oscillations in $G_{DQD}^{(2)}(\phi)$, which is defined as
\begin{eqnarray}
\Delta G_{DQD}^{(2)}\equiv G_{DQD}^{(2)}(\phi_{{\scriptsize \mbox{max}}}),
\end{eqnarray}
where $\phi_{max}$ is the phase when $G_{DQD}^{(2)}(\phi)$ is maximal in the AB oscillation. When $\Gamma_{\varphi}$ increases, $\Delta G_{DQD}^{(2)}$ has a peak as shown in Fig. \ref{NLC-GaP-alpha} (b) when $\epsilon_d/\hbar\Gamma=1$ and $t_c/\hbar\Gamma=-1$. To understand the behavior of $G_{DQD}^{(2)}(\phi)$ in the weak coupling regime ($\Gamma_{\varphi}\ll\Gamma$) in Fig. \ref{NLC-GaP-alpha} (b), we consider the situation where $\epsilon_d=-t_c$, and $|\alpha|\ll 1$. Then, we obtain
\begin{eqnarray}
\left.\frac{\partial\Delta G_{DQD}^{(2)}}{\partial\Gamma_{\varphi}}\right|_{\Gamma_{\varphi}=0}=\frac{32{t_c}^2|\sin(\phi_{{\scriptsize \mbox{max}}})|}{128{t_c}^4+24{t_c}^2(\hbar\Gamma)^2+(\hbar\Gamma)^4}|\alpha|.
\end{eqnarray}
Up to the order of $|\alpha|$, we have
\begin{eqnarray}
\phi_{{\scriptsize \mbox{max}}}= \left\{\begin{array}{cc}
    \frac{\pi}{2} & (\alpha>0) \\ 
    -\frac{\pi}{2} & (\alpha<0) \\ 
  \end{array}
\right..
\end{eqnarray}
Thus, the slope of $\Delta G_{DQD}^{(2)}$ for $\Gamma_{\varphi}\ll\Gamma$ increases with $|\alpha|$.

In the limit of $\Gamma_{\varphi}\gg\Gamma$, the amplitude of AB oscillations in the lowest-order nonlinear conductance coefficient has the following leading term of an asymptotic series for $|\alpha|\neq 1$
\begin{eqnarray}
\Delta G_{DQD}^{(2)}\sim\frac{e^3}{h}\frac{32(\hbar\Gamma)^2\left| t_c\alpha\left[\epsilon_d-t_c\alpha\cos(\phi_{{\scriptsize \mbox{max}}}) \right]\sin(\phi_{{\scriptsize \mbox{max}}}) \right|}{(1-\alpha^2)^3}\left(\frac{1}{\hbar\Gamma_{\varphi}}\right)^5.
\end{eqnarray}
When $\epsilon_d=-t_c$ and $|\alpha|\ll 1$, we have
\begin{eqnarray}
\Delta G_{DQD}^{(2)}\sim\frac{e^3}{h}\frac{32(\hbar\Gamma)^2{t_c}^2|\alpha|}{(\hbar\Gamma_{\varphi})^5}.
\end{eqnarray}
Under the condition where $|\alpha|\ll 1$ is not satisfied, it is difficult to discuss the asymptotic behavior of $\Delta G_{DQD}^{(2)}$ since $\phi_{{\scriptsize \mbox{max}}}$ is a function of $\alpha$ and $\Gamma_{\varphi}$. In general, $\Delta G_{DQD}^{(2)}$ is a function of $|\alpha|$ since $G_{DQD}^{(2)}$ is invariant under the transformation that we change the sign of $\alpha$ and shift the phase by $\pi$. Moreover, we plot the $|\alpha|$ dependence of the peak height of $\Delta G_{DQD}^{(2)}$ (indicated as $\Delta G_{DQD,{\scriptsize \mbox{max}}}^{(2)}$) as shown in Fig. \ref{NLC-GaP-alpha}(c). The peak height increases monotonically as $|\alpha|$ increases in the same way the visibility of the AB oscillation in $G_{DQD}^{(1)}(\phi)$.

\section{Summary\label{summary}}
We studied linear and nonlinear three-terminal transport through an AB interferometer containing a DQD using the nonequilibrium Green's function method. We introduced coherent indirect coupling between two quantum dots via a voltage probe $\varphi$. The linear conductance exhibits phase symmetry without depending on the various parameters of the model. However, in the nonlinear transport regime, the phase symmetry is broken and the phase of the AB oscillations shifts. We showed that the lowest-order nonlinear conductance coefficient with respect to the bias voltage contributes to the phaseshift. In particular, when $|\alpha|\hbar\Gamma_{\varphi}, |t_c|\ll\hbar\Gamma$, where we can neglect the higher harmonic components of the AB oscillations, we proved that the sign of the lowest-order nonlinear conductance coefficient is directly related to that of the phaseshift, and the value of the phaseshift is determined by the quotient between the linear conductance and the lowest-order nonlinear conductance coefficient. In the weak coherent indirect coupling and low bias voltage regimes, the phaseshift is independent of the coherent indirect coupling parameter and monotonically increasing function with respect to the source-drain bias voltage. Moreover, we obtained a condition where the direction the phase of the AB oscillation shifts from the lowest-order nonlinear conductance coefficient. In the coupling $\Gamma_{\varphi}$ dependence of the visibility of the AB oscillations in the linear conductance, we found that the visibility has a maximum value except when $\alpha=1$. When $\alpha=1$, the visibility is a monotonically increasing function of $\Gamma_{\varphi}$.

\begin{acknowledgments}
We thank Yuli V. Nazarov, S. Tarucha, Y. Utsumi, T. Hatano, S. Amaha, and S. Sasaki for useful discussions and valuable comments. Part of this work is supported financially by JSPS MEXT Grant-in-Aid for Scientific Research on Innovative Areas (21102003) and Funding Program for World-Leading Innovative R\&D Science and Technology (FIRST).
\end{acknowledgments}

\appendix

\section{Derivation of probability conservation\label{probability-condition}}
In this appendix, we derive the relation of the probability conservation given by Eq. (\ref{aps1}). The left-hand-side of Eq. (\ref{aps1}) is
\begin{eqnarray}
\sum_{\xi\neq\nu}T_{\nu\xi}(\epsilon,\phi)&=&\sum_{\xi\neq\nu}\mbox{Tr}\left\{\bm{G}^r(\epsilon,\phi)\bm{\Gamma}^{\xi}(\phi)\bm{G}^a(\epsilon,\phi)\bm{\Gamma}^{\nu}(\phi) \right\}\nonumber\\
&=&\mbox{Tr}\left\{\bm{G}^r(\epsilon,\phi)\bm{\Gamma}(\phi)\bm{G}^a(\epsilon,\phi)\bm{\Gamma}^{\nu}(\phi) \right\}-\mbox{Tr}\left\{\bm{G}^r(\epsilon,\phi)\bm{\Gamma}^{\nu}(\phi)\bm{G}^a(\epsilon,\phi)\bm{\Gamma}^{\nu}(\phi) \right\}.\label{appen1}
\end{eqnarray}
Here the Dyson's equation for the retarded Green's function is
\begin{eqnarray}
\bm{G}^r(\epsilon,\phi)&=&\bm{g}^r(\epsilon)+\bm{g}^r(\epsilon)\bm{\Sigma}^r(\phi)\bm{G}^r(\epsilon,\phi)\nonumber\\
&=&\bm{g}^r(\epsilon)-\frac{i}{2}\bm{g}^r(\epsilon)\bm{\Gamma}(\phi)\bm{G}^r(\epsilon,\phi),
\end{eqnarray}
where $\bm{g}^r(\epsilon)$ is the retarded Green's function of an isolated DQD. Thus, we have
\begin{eqnarray}
\bm{\Gamma}(\phi)&=&i\left\{\bm{\Sigma}^r(\phi)-\bm{\Sigma}^a(\phi) \right\}\nonumber\\
&=&i\left\{[\bm{G}^a(\epsilon,\phi)]^{-1}-[\bm{G}^r(\epsilon,\phi)]^{-1} \right\}.
\end{eqnarray}
Therefore, Eq. (\ref{appen1}) is
\begin{eqnarray}
\sum_{\xi\neq\nu}T_{\nu\xi}(\epsilon,\phi)&=&i\mbox{Tr}\left\{\bm{G}^r(\epsilon,\phi)\left[[\bm{G}^a(\epsilon,\phi)]^{-1}-[\bm{G}^r(\epsilon,\phi)]^{-1} \right]\bm{G}^a(\epsilon,\phi)\bm{\Gamma}^{\nu}(\phi) \right\}-\mbox{Tr}\left\{\bm{G}^r(\epsilon,\phi)\bm{\Gamma}^{\nu}(\phi)\bm{G}^a(\epsilon,\phi)\bm{\Gamma}^{\nu}(\phi) \right\}\nonumber\\
&=&i\mbox{Tr}\left\{\bm{G}^r(\epsilon,\phi)\bm{\Gamma}^{\nu}(\phi)-\bm{G}^a(\epsilon,\phi)\bm{\Gamma}^{\nu}(\phi) \right\}-\mbox{Tr}\left\{\bm{G}^r(\epsilon,\phi)\bm{\Gamma}^{\nu}(\phi)\bm{G}^a(\epsilon,\phi)\bm{\Gamma}^{\nu}(\phi) \right\}\nonumber\\
&=&i\mbox{Tr}\left\{\bm{\Gamma}^{\nu}(\phi)\bm{G}^r(\epsilon,\phi)-\bm{\Gamma}^{\nu}(\phi)\bm{G}^a(\epsilon,\phi) \right\}-\mbox{Tr}\left\{\bm{G}^r(\epsilon,\phi)\bm{\Gamma}^{\nu}(\phi)\bm{G}^a(\epsilon,\phi)\bm{\Gamma}^{\nu}(\phi) \right\}\nonumber\\
&=&i\mbox{Tr}\left\{\bm{\Gamma}^{\nu}(\phi)\bm{G}^a(\epsilon,\phi)\left([\bm{G}^a(\epsilon,\phi)]^{-1}-[\bm{G}^r(\epsilon,\phi)]^{-1} \right)\bm{G}^r(\epsilon,\phi) \right\}-\mbox{Tr}\left\{\bm{G}^r(\epsilon,\phi)\bm{\Gamma}^{\nu}(\phi)\bm{G}^a(\epsilon,\phi)\bm{\Gamma}^{\nu}(\phi) \right\}\nonumber\\
&=&\mbox{Tr}\left\{\bm{\Gamma}^{\nu}(\phi)\bm{G}^a(\epsilon,\phi)\bm{\Gamma}(\phi)\bm{G}^r(\epsilon,\phi) \right\}-\mbox{Tr}\left\{\bm{G}^r(\epsilon,\phi)\bm{\Gamma}^{\nu}(\phi)\bm{G}^a(\epsilon,\phi)\bm{\Gamma}^{\nu}(\phi) \right\}\nonumber\\
&=&\sum_{\xi\neq\nu}\mbox{Tr}\left\{\bm{G}^r(\epsilon,\phi)\bm{\Gamma}^{\nu}(\phi)\bm{G}^a(\epsilon,\phi)\bm{\Gamma}^{\xi}(\phi) \right\}\nonumber\\
&=&\sum_{\xi\neq\nu}T_{\xi\nu}(\epsilon,\phi).
\end{eqnarray}

\section{Derivation of phase symmetry\label{derivation-phase}}
In this appendix, we prove the phase symmetry relation (\ref{phase-symmetry}). Using Eqs. (\ref{linear-conductance}) and (\ref{trs}), the linear conductance is
\begin{eqnarray}
G_{DQD}^{(1)}(-\phi)&=&\frac{e^2}{h}\frac{T_{SD}(\epsilon=0,\phi)T_{\varphi S}(\epsilon=0,\phi)+T_{SD}(\epsilon=0,\phi)T_{\varphi D}(\epsilon=0,\phi)+T_{S\varphi}(\epsilon=0,\phi)T_{\varphi D}(\epsilon=0,\phi)}{T_{\varphi S}(\epsilon=0,\phi)+T_{\varphi D}(\epsilon=0,\phi)}.\label{aps2}
\end{eqnarray}
Using relation (\ref{aps1}), the numerator of the right-hand-side of Eq. (\ref{aps2}) can be rewritten as
\begin{eqnarray}
&&T_{SD}(\epsilon=0,\phi)T_{\varphi S}(\epsilon=0,\phi)+\left[T_{SD}(\epsilon=0,\phi)+T_{S\varphi}(\epsilon=0,\phi) \right]T_{\varphi D}(\epsilon=0,\phi)\nonumber\\
&=&T_{SD}(\epsilon=0,\phi)T_{\varphi S}(\epsilon=0,\phi)+\left[T_{DS}(\epsilon=0,\phi)+T_{\varphi S}(\epsilon=0,\phi) \right]T_{\varphi D}(\epsilon=0,\phi).\label{aps3}
\end{eqnarray}
By continuous use of Eq. (\ref{aps1}), Eq. (\ref{aps3}) is
\begin{eqnarray}
&&\left[T_{SD}(\epsilon=0,\phi)+T_{\varphi D}(\epsilon=0,\phi) \right]T_{\varphi S}(\epsilon=0,\phi)+T_{DS}(\epsilon=0,\phi)T_{\varphi D}(\epsilon=0,\phi)\nonumber\\
&=&\left[T_{DS}(\epsilon=0,\phi)+T_{D\varphi}(\epsilon=0,\phi) \right]T_{\varphi S}(\epsilon=0,\phi)+T_{DS}(\epsilon=0,\phi)T_{\varphi D}(\epsilon=0,\phi)\nonumber\\
&=&T_{DS}(\epsilon=0,\phi)\left[T_{\varphi S}(\epsilon=0,\phi)+T_{\varphi D}(\epsilon=0,\phi) \right]+T_{\varphi S}(\epsilon=0,\phi)T_{D\varphi}(\epsilon=0,\phi).\label{aps4}
\end{eqnarray}
Therefore, by comparison with Eq. (\ref{linear-conductance}), we obtain the phase symmetry relation
\begin{eqnarray}
G_{DQD}^{(1)}(-\phi)&=&G_{DQD}^{(1)}(\phi).
\end{eqnarray}

\section{Derivation of antisymmetricity for lowest-order nonlinear conductance coefficient\label{anti-ap}}
Here we show that the lowest-order nonlinear conductance coefficient is asymmetric with respect to the flux. Using relations (\ref{trs}) and (\ref{mirror}), when the direction of the magnetic flux is reversed, the flux-dependent contribution in the 1st line of the right-hand side in Eq. (\ref{nlc}) is
\begin{eqnarray}
\left\{\frac{T_{S\varphi}(\epsilon=0,-\phi)-T_{D\varphi}(\epsilon=0,-\phi)}{T_{S\varphi}(\epsilon=0,-\phi)+T_{D\varphi}(\epsilon=0,-\phi)} \right\}^2=\left\{\frac{T_{D\varphi}(\epsilon=0,\phi)-T_{S\varphi}(\epsilon=0,\phi)}{T_{D\varphi}(\epsilon=0,\phi)+T_{S\varphi}(\epsilon=0,\phi)} \right\}^2.\label{first}
\end{eqnarray}
Then, this contribution is symmetric with respect to $\phi$. Similarly, the 2nd line of the right-hand side in Eq. (\ref{nlc}) is
\begin{eqnarray}
&&\left.\frac{\partial T_{\varphi S}(\epsilon,-\phi)}{\partial\epsilon}\right|_{\epsilon=0}-\frac{\left. \frac{\partial T_{S\varphi}(\epsilon,-\phi)}{\partial\epsilon}\right|_{\epsilon=0}+\left.\frac{\partial T_{D\varphi}(\epsilon,-\phi)}{\partial\epsilon}\right|_{\epsilon=0}}{T_{S\varphi}(\epsilon=0,-\phi)+T_{D\varphi}(\epsilon=0,-\phi)}T_{\varphi S}(\epsilon=0,-\phi)\nonumber\\
&=&\left.\frac{\partial T_{S\varphi}(\epsilon,\phi)}{\partial\epsilon}\right|_{\epsilon=0}-\frac{\left. \frac{\partial T_{\varphi S}(\epsilon,\phi)}{\partial\epsilon}\right|_{\epsilon=0}+\left.\frac{\partial T_{S\varphi}(\epsilon,\phi)}{\partial\epsilon}\right|_{\epsilon=0}}{T_{\varphi S}(\epsilon=0,\phi)+T_{S\varphi}(\epsilon=0,\phi)}T_{S\varphi}(\epsilon=0,\phi)\nonumber\\
&=&\frac{\left. \frac{\partial T_{S\varphi}(\epsilon,\phi)}{\partial\epsilon}\right|_{\epsilon=0}T_{\varphi S}(\epsilon=0,\phi)-\left.\frac{\partial T_{\varphi S}(\epsilon,\phi)}{\partial\epsilon}\right|_{\epsilon=0}T_{S\varphi}(\epsilon=0,\phi)}{T_{\varphi S}(\epsilon=0,\phi)+T_{S\varphi}(\epsilon=0,\phi)}\nonumber\\
&=&\frac{-\left.\frac{\partial T_{\varphi S}(\epsilon,\phi)}{\partial\epsilon}\right|_{\epsilon=0}\left[T_{S\varphi}(\epsilon=0,\phi)+T_{D\varphi}(\epsilon=0,\phi) \right]+\left[\left.\frac{\partial T_{S\varphi}(\epsilon,\phi)}{\partial\epsilon}\right|_{\epsilon=0}+\left.\frac{\partial T_{D\varphi}(\epsilon,\phi)}{\partial\epsilon}\right|_{\epsilon=0} \right] T_{\varphi S}(\epsilon=0,\phi)}{T_{D\varphi}(\epsilon=0,\phi)+T_{S\varphi}(\epsilon=0,\phi)}\nonumber\\
&=&-\left[\left.\frac{\partial T_{\varphi S}(\epsilon,\phi)}{\partial\epsilon}\right|_{\epsilon=0}-\frac{\left.\frac{\partial T_{S\varphi}(\epsilon,\phi)}{\partial\epsilon}\right|_{\epsilon=0}+\left.\frac{\partial T_{D\varphi}(\epsilon,\phi)}{\partial\epsilon}\right|_{\epsilon=0}}{T_{S\varphi}(\epsilon=0,\phi)+T_{D\varphi}(\epsilon=0,\phi)}T_{\varphi S}(\epsilon=0,\phi) \right].\label{second}
\end{eqnarray}
Then, this contribution is asymmetric with respect to $\phi$. As a result, from Eqs. (\ref{nlc}), (\ref{first}) and (\ref{second}), we find that $G_{DQD}^{(2)}(-\phi)=-G_{DQD}^{(2)}(\phi)$.

\section{Derivation of visibility (\ref{visibility-limit})\label{ap1}}
In this appendix, we derive expression (\ref{visibility-limit}) for the visibility of the AB oscillations in the linear conductance in the limit of $\alpha=1$ and $\Gamma_{\varphi}\to\infty$. We consider the mirror symmetry ($\epsilon_1=\epsilon_2=\epsilon_d$, $\Gamma_S=\Gamma_D=\Gamma$, and $\Gamma_{\varphi1}=\Gamma_{\varphi2}=\Gamma_{\varphi}$). Under this condition, the linear conductance is given by
\begin{eqnarray}
G_{DQD}^{(1)}(\phi)&=&\frac{e^2}{h}T_{DS}(\epsilon=0,\phi)\nonumber\\
&=&\frac{e^2}{h}\frac{(\hbar\Gamma)^2}{4(\epsilon_d-t_c\cos\phi)^2+(\hbar\Gamma)^2},
\end{eqnarray}
since, in the limit of $\alpha=1$ and $\Gamma_{\varphi}\to\infty$, we have
\begin{eqnarray}
T_{\varphi S}(\epsilon=0,\phi)=T_{D\varphi}(\epsilon=0,\phi)\propto\frac{\Gamma}{\Gamma_{\varphi}},
\end{eqnarray}
and thus the second term in Eq. (\ref{linear-conductance}) vanishes. Then, from the following condition
\begin{eqnarray}
\frac{\partial G_{DQD}^{(1)}(\phi)}{\partial\phi}=0,
\end{eqnarray}
we have the condition of $\phi$ for extreme values:
\begin{eqnarray}
\sin\phi=0\ ,\ \cos\phi=\frac{\epsilon_d}{t_c}.
\end{eqnarray}
If $\epsilon_d\neq|t_c|$, the condition $\cos\phi=\epsilon_d/t_c$ leads to $\phi\neq n\pi$, where $n$ is an integer. Then, we have
\begin{eqnarray}
G_{DQD,max}^{(1)}=\frac{e^2}{h}\ ,\ G_{DQD,min}^{(1)}=\min\left(G_{DQD}^{(1)}(\phi=0),G_{DQD}^{(1)}(\phi=\pi) \right).
\end{eqnarray}
Similarly, for $\epsilon_d=|t_c|$, the condition $\cos\phi=\epsilon_d/t_c$ leads to $\phi=0$ or $\pi$. Then, we have
\begin{eqnarray}
G_{DQD,max}^{(1)}=\frac{e^2}{h}\ ,\ G_{DQD,min}^{(1)}=\min\left(G_{DQD}^{(1)}(\phi=0),G_{DQD}^{(1)}(\phi=\pi) \right).
\end{eqnarray}
Therefore, from the definition (\ref{visibility-definition}), we obtain the expression of visibility (\ref{visibility-limit}).

\end{document}